# Atomic-scale Mapping Unravel Anisotropic Dissolution Behaviors of Gibbsite Nanosheets

*Xiaoxu Li[a,#], Qing Guo[b,#], Yatong Zhao[a,#], Ping Chen[a], Benjamin A Legg[a], Lili Liu[a], Chang Liu[c], Qian Chen[c], Zheming Wang[a], James J. De Yoreo[a], Carolyn I Pearce[a], Aurora E. Clark[a,b,\*], Kevin M. Rosso[a,\*], and Xin Zhang[a,\*],*

[a] *Physical and Computational Sciences Directorate, Pacific Northwest National Laboratory, Richland, Washington 99354, United States*

[b] *Department of Chemistry, University of Utah, Salt Lake City, Utah 84112, United States*

[c] *Department of Materials Science and Engineering, Materials Research Laboratory, Beckman Institute for Advanced Science and Technology, University of Illinois at Urbana−Champaign, Urbana, Illinois 61801, United States*

[#] These authors contributed equally to this work

**Corresponding Author:** xin.zhang@pnnl.gov (X.Z.), aurora.clark@utah.edu (A.E.C.), and kevin.rosso@pnnl.gov (K.M.R.)

**Abstract**

This study examines the anisotropic dissolution of the basal plane gibbsite (γ-Al(OH)$_3$) nanoplates in sodium hydroxide solution using *in situ* atomic force microscopy (AFM) and density functional theory (DFT) calculations. In the surface-reaction controlled regime, *in situ* AFM measurements reveal anisotropic dissolution of hillocks and etch pits on the gibbsite basal plane, with preferred dissolution directions alternating between layers. The mirror-symmetric pattern of dissolution preference between adjacent gibbsite aluminum hydroxide sheet, observed along the crystallographic a-c plane, results from the matching symmetry between the structures of the adjacent (001) and (002) crystal planes. Consequently, the overall dissolution rate of gibbsite nanoplates exhibits crystallographic a-c plane symmetry, as the rate of parallel steps is governed by the slower ones. DFT calculations suggest that the anisotropic dissolution is partially due to the orientation and strength of Al-OH-Al linkages pair within gibbsite surface structure. These findings offer a comprehensive understanding of anisotropic dissolution behavior of gibbsite and illuminate the mechanisms behind preferential dissolution.

**Introduction**

Gibbsite (γ-Al(OH)$_3$) is one of the most common aluminum-bearing minerals and is found in a wide range of geological environments, including weathered rocks, laterite deposits, and soils[1-4]. The study of gibbsite has gained significant attention due to their simple chemistry and structure, making it an ideal model mineral in surface chemistry[5, 6], nucleation[7-9], and mineral dissolution[10-15]. The knowledge of reactions occurring at the gibbsite-solution interface provides a foundation for understanding the reactions of all minerals that contain aluminum in octahedral coordination.

Mineral dissolution contributes to the process controlling global geochemical cycling[16], lithosphere evolution[17], the quality and quantity of groundwater and surface water[18], the formation, exploration, and exploitation of mineral resources. The dissolution of gibbsite under alkaline conditions is of particular importance as it can impact various industrial and environmental processes. One of the most important uses of gibbsite is in the production of aluminum metal[19]. The Bayer process, which involves the dissolution of gibbsite in sodium hydroxide solution, is the most widely used method for extracting aluminum from bauxite ores[2]. The similar process is used to treat high-level radioactive waste at the Hanford site in Washington, USA[20-23]. A detailed understanding the dissolution behavior of gibbsite in alkaline solutions, such as sodium hydroxide (NaOH), is essential for optimizing industrial processes and developing new applications, including alumina production, soil remediation, and waste management[21].

During the dissolution of gibbsite in an alkaline solution, the aluminum ion undergoes a transformation from a aluminum octahedral structure in crystal to a tetrahedral ion, according to the overall reaction: $Al(OH)_3 + OH^- = Al(OH)_4^-$. Inferred from ion release rates, various dissolution models for gibbsite in alkaline solutions have been proposed, based on different factors, including NaOH concentration, temperature, and gibbsite morphology and surface area[24-30]. However, a universally accepted mechanism that encompasses all experimental observations is still lacking. For example, according to dissolution kinetics, the reported reaction order of gibbsite dissolution in alkaline conditions with respect to sodium hydroxide concentration ranges from 0.77 to 2[24, 31] [25] [30] [27]. Although these studies derived useful empirical rate laws, the role of atomic structure, bonding, and surface morphology were not incorporated in the rate equation. These highlight the need for further research to elucidate the specific surface reactions and processes governing gibbsite dissolution in alkaline conditions at microscale. Recent advances in atomic force microscopy (AFM) provide a promising approach to address this issue. High-speed AFM has the capability to capture images of interfacial surface structure at near atomic-level resolution, providing the opportunity to directly observe the step movement in dissolution. To the best of our knowledge, only a few studies have reported the investigation of aluminum hydroxides dissolution behavior using AFM[32] or other *in situ* high spatiotemporal resolution characterization techniques (e.g., liquid cell TEM[33, 34]), and these studies have been limited to acidic solution or irradiated condition.

In this study, we employ *in situ* AFM to reveal the dissolution behavior of gibbsite basal plane in NaOH solutions with concentrations ranging from 0.1 M to 0.5 M. Our results reveal that the dissolution rate of gibbsite nanoplates varies significantly with crystal orientation and changes layer by layer. Based on DFT calculation, the anisotropic dissolution behavior is attributed to the different strengths of Al-OH-Al pair linkages connecting adjacent aluminum atoms in different crystallographic directions. Our findings provide insights into the fundamental understanding of the dissolution behavior of gibbsite and other aluminum-containing minerals in alkaline environments, which have important implications in development of dissolution model.

**Results and discussion**

### Morphology of as-synthesized gibbsite nanoplates

Well-crystalline and uniform gibbsite nanoplates were synthesized according to previously established synthesis methodology[35, 36]. The X-ray diffraction pattern confirmed the pure gibbsite phase of the synthesized material (Figure S1). In order to establish the link between crystal morphology and structure, and to aid in determining the crystallographic orientation of gibbsite nanoplates in the in situ AFM measurements, we characterized the synthesized gibbsite using

TEM. This step is essential, as we found significant anisotropy in the dissolution of the gibbsite basal plane, as discussed further below.

Figure 1a illustrates the bright field TEM image of as-synthesized gibbsite, which appears as pseudohexagonal-shaped plates elongated along a specific direction. The SEM images showed the longest dimension of nanoplates is within the range of 250 to 550 nm (Figure S2). Based on the reported gibbsite crystal structure[37], the selective area electron diffraction pattern (SAED, Figure 1c) confirm that the preferential crystal growth direction of gibbsite is [100]. Additional gibbsite nanoplates were also indexed using SAED, consistently demonstrating that the gibbsite extends along the [100] direction (Figure S3). The HRTEM images (Figure 1b) further indicate that the longer four edge faces are composed of {110} crystal facets. The thickness of the gibbsite nanoplates, estimated to be approximately 25 nm, was observed by viewing a bright field TEM image of the gibbsite nanoplates parallel along the gibbsite basal [010] zone axis (Figure 1d and f). The HRTEM image reveals that one of the nanoplate side faces is $(20\bar{1})$, forming an acute dihedral angle of 70° with (001) face, and therefore $(\bar{2}01)$ face on the other side have obtuse dihedral angle of 110° with (001) face. Due to the very similar interplanar spacing between gibbsite (110) and (002) crystal planes (0.4380 vs. 0.4328 nm), it is difficult to distinguish them accurately in high-resolution AFM images. Hence, we can determine them based on the extension direction of the crystal basal plane. Additionally, we can further determine the <100> direction by observing the dihedral angle at the edge of the (001) face in the AFM image.

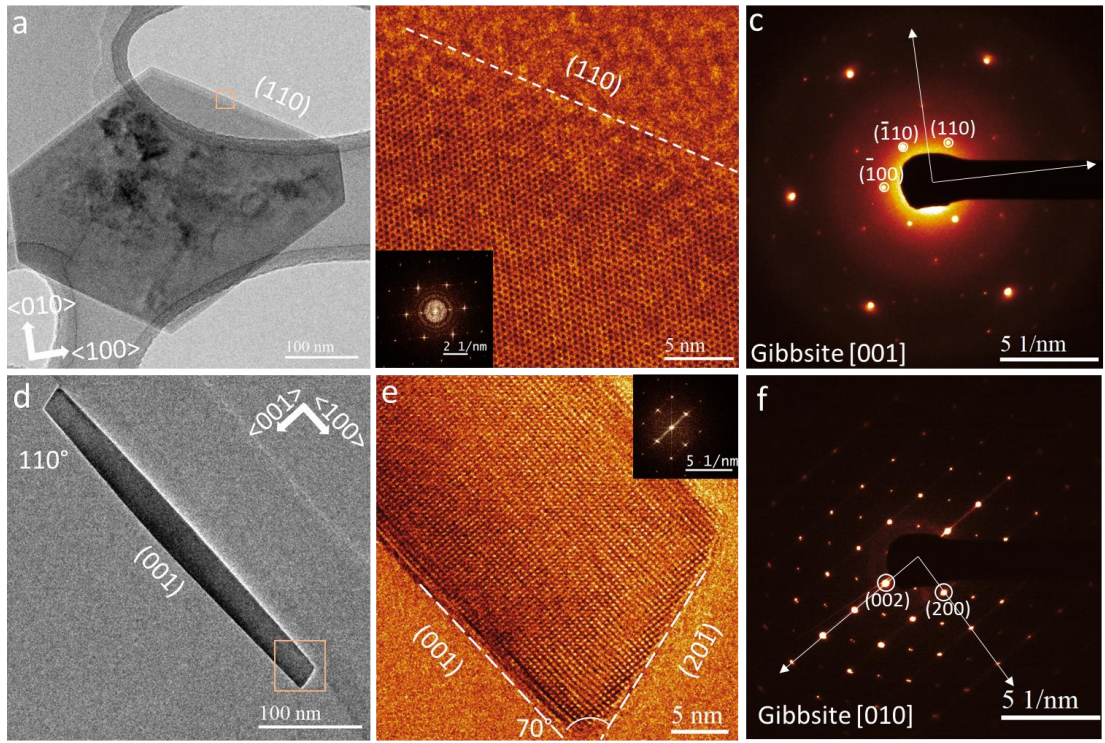

Figure 1. The bright field TEM image (a), HRTEM image (b), SAED pattern (c) of gibbsite nanoplate along zone axis [001], and the bright filed TEM image (d), HRTEM image (e), and SAED pattern (f) of gibbsite nanoplates along zone axis [010].

**Anisotropic dissolution of hillock on gibbsite basal plane**

As shown in AFM phase image at the beginning of dissolution (Figure 2a), growth hillock was observed as the advancement of 5 Å high single aluminum hydroxides steps separated by atomically flat terraces. These layers were generated as steps in a continuous mode from a double screw dislocation with Burgers vector *b* of one gibbsite unite cell (two aluminum hydroxide layer) at the center of (001) surface. Generally, hillocks exhibited a pseudohexagonal geometry with six flanks compose of {110} and {100} step on the basal plane as labeled in Figure 2a. Figure 2d and 2e plot the spirals counterplot on the top of hillock. In the beginning of dissolution, the spirals near the dislocation exhibit rounded parallelogram geometries compose of {110} step, such that symmetrically related step are paired to from two nonequivalent sides in the synthesis condition (near natural-pH condition). The asymmetric parallelogram but not symmetric rhombic geometry of these spirals also indicates the anisotropic growth rate of four {110} step during the synthesis. Interestingly, the geometries of the two spirals formed by a double screw dislocation are not identical, but they exhibit a mirror symmetry respect to the crystallographic a-c plane (h0k). The same symmetry is also reflected in the atomic structure of adjacent layers of aluminum hydroxide octahedra (gibbsite (001) and (002)) as shown in Figure 3. Previous research has commonly assumed that the hexagonal geometry of the basal plane of gibbsite confers identical reactivity in four {110} steps[38]. However, the morphology of gibbsite hillock suggests that different orientated steps have different intrinsic reactivity and might lead the anisotropic dissolution in alkaline condition as shown below.

In 0.1M NaOH, the *in situ* AFM measurements indicate that dissolution occurs through the retreat of surface steps rather than the nucleation of etch pits at screw dislocations. The evolution of two spirals over a 162-minute period is shown in Figure 2d and 2e, with the position of the spiral contour line corrected by the center of the screw dislocation to eliminate image drift. There are two different types of steps on the gibbsite basal plane (as shown in Figure 6d). One is "zigzag" type, which has six different orientations: four {110} and two {100} steps on the basal plane. The other one is "armchair" type, which also has six different orientations: four {310} and two {010} step on the basal plane[32]. The retreat rates of differently oriented steps were measured based on the slope of the linear fit of the retreat distance of the step along the normal direction over time. To avoid the effects of the stress field around the dislocation center, only the dissolution rate of the outermost steps located far away from the dislocation were measured.

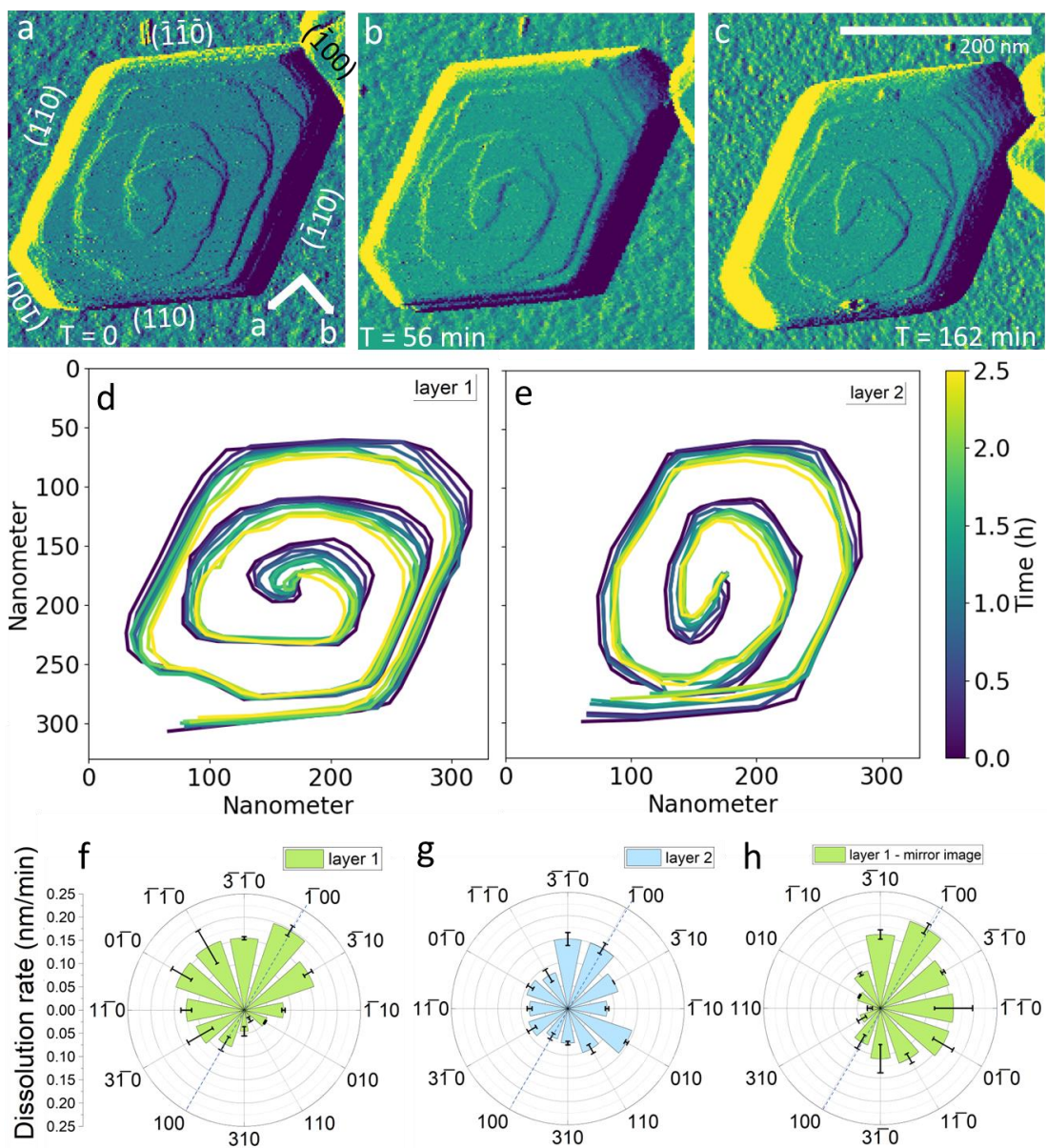

Figure 2. *In situ* AFM amplitude images of gibbsite nanoplates dissolving in 0.1 M NaOH. (d and e) The evolution of two spirals originated from same double screw dislocation at gibbsite basal plane. (f) The step velocity (i.e., dissolution rate) as a function of step orientation of the spiral in (d). (g) The step velocity (i.e., dissolution rate) as a function of step orientation of the spiral in (e). (h) The mirror image of orientation-dependent step velocity of spiral in (d) respect to a-c plane.

Figure 2f displays the dissolution rates of 12 zigzag- and armchair-type steps of layer 1, as shown in Figure 2d, layer 1 exhibits notable dissolution anisotropy, with the ($\bar{1}$00) step showing the highest dissolution rate (0.197 ± 0.012 nm/min), which is almost one order of magnitude faster than the slowest step ($\bar{1}\bar{1}0$) (0.023 ± 0.005 nm/mins). As a result, the terrace width is larger in the ($\bar{1}00$) direction compared to other directions, as shown in Figure 2c (t = 162 minutes). Furthermore, the dissolution rate of steps on crystal surfaces is not the same for two opposite orientations. The dissolution rate for steps oriented in the ($\bar{1}00$), ($\bar{3}\bar{1}0$), ($\bar{1}\bar{1}0$), and ($0\bar{1}0$) compared to is significantly greater their oppositely oriented counterparts.

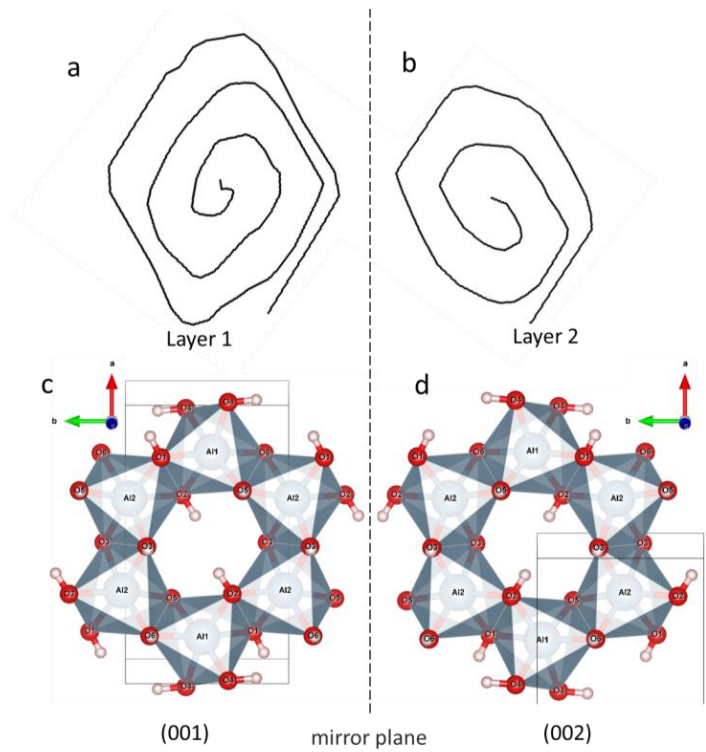

Figure 3. Schematic diagram of the plane symmetry in the crystallography a-c plane for the spirals of adjacent layers of gibbsite basal plane (a and b) and the adjacent (001) and (002) crystal structures (c and d).

As shown in Figure 2g, for the layer 2 which is located below layer 1, the anisotropy of its dissolution is also significant, but the preferred dissolution direction is different. Like layer 1, zigzag-type ($\bar{1}00$) step and adjacent armchair-type ($\bar{3}\bar{1}0$) and ($\bar{3}10$) steps on the layer 2 have higher dissolution rate. However, in contrast to the layer 1, the (010) step is significantly higher than opposite (0$\bar{1}$0). The differences in anisotropic dissolution of adjacent gibbsite aluminum hydroxide octahedral layers corroborate their crystallographic symmetry as shown in Figure 3c and 3d. For ease of comparison, Figure 2h illustrates the dissolution rate spectra of layer 1 after mirror symmetry along the crystallographic ac-plane (shown in dash line). Figures 2g and 2h exhibit similar anisotropic dissolving trends, but with less pronounced anisotropy. This is partly attributable to the influence of the upper layer 1. The slower dissolution rate of layer 1 on the ($\bar{1}$10) step limits the rate of step below, even though the crystal structure correlation predicts faster dissolution for the ($\bar{1}$10) step of layer 2 compared to (1$\bar{1}$0).

**Anisotropic growth of etch pits at gibbsite basal plane**

In 0.2 M NaOH, dissolution occurs by the creation of pits and the retreat of steps at the surface (Figure 4). The sample had already been mixed with NaOH solution for 23 minutes prior to image first slice as shown in figure 4a. The crystallographic orientation of gibbsite nanoplates has been confirmed by a high-resolution AFM image (Figure 4c) and indexing of the corresponding FFT image (Figure 4d), which was obtained after the last slice (Figure 4b) of particle dissolution measurement.

There are two types of pits on the surface. The first type is "shallow" pit consisted of single layer as shown by red dash. The second type is "deep" pits, which are 4-5 atomic layers deep as shown by white dash line. It is useful to note that theses "deep" pits have slop of sides of only 3-6°. Thus they are deep compared to single-layer pits, but still shallow or gradual on a general scale[39]. As shown in the video S1, "shallow" pits are the dominate features during dissolution. They occurred frequently and randomly but were short lived, showing that "shallow" pits were initiated by thermal events or random point defects at the surface[39]. The depth of the pits is mostly ~5 Å, corresponding to a depth of one aluminum hydroxide layer. Pits merged with each other resulting in diamond shape or complex shape and were mostly consisted of {110} steps.

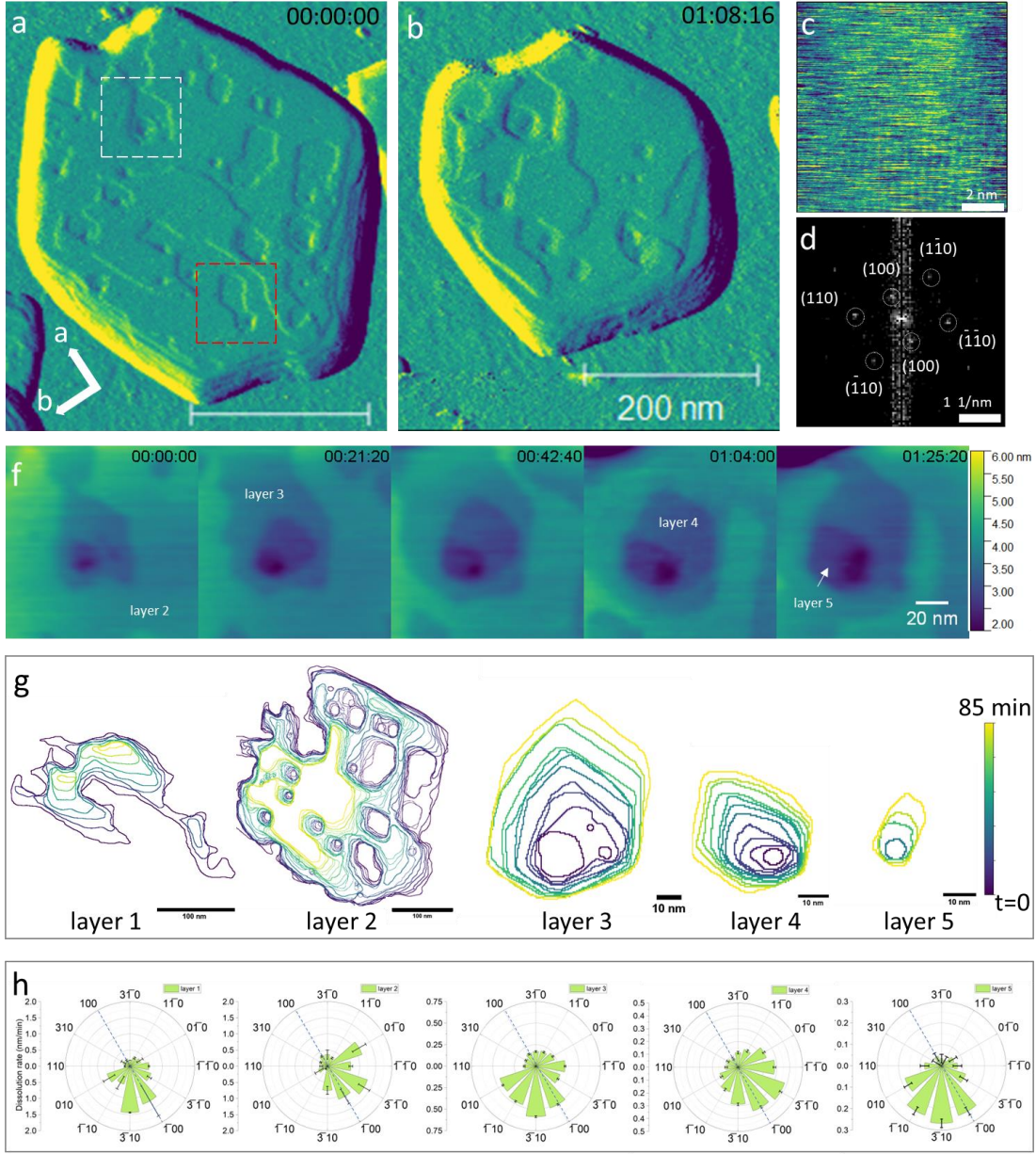

Figure 4. (a and b) *In situ* AFM amplitude images of gibbsite dissolving in 0.2 M NaOH. The red and white dash lines show the "shallow" and "deep" etch pit. (c) The high-resolution AFM image of the nanoplates in (b). (d) FFT of (c) and indexes show the crystal orientation. (f) The AFM height image of evolution of "deep" etch pits. (g) Contour plots of step edge over time of the different layers at basal plane, in which layers 3 to 5 represent the contours of different layers of the deep pit shown in (f). The step velocities (i.e., dissolution rate) of layers shown in (g) as function of step orientation.

"Deep" pits occurred less frequently. As shown in Figure 4e, the process of a "deep" pit dissolution is illustrated in the magnified AFM height images, which are taken from the area enclosed by the white box in Figure 4a. The step heights on both sides of the deep pit are the same, indicating that the center was not originally a screw dislocation. What's more, the nucleation of etch pit didn't occurred at the screw dislocation previously exist in nanoplate in even higher NaOH concentration (0.4 M) as shown in figure S4. Furthermore, we observed that these deep pits did not continue to extend further after expanding for 3-4 layers. Instead, the bottom became flat, as shown in Figure S5. Brown dissolved gibbsite in a far-from-equilibrium (5M) NaOH solution, like the type used in the aluminum industry, and observed that the etch pit was shallow with a flat bottom. This morphology was similar to what we observed in our SEM analysis[40]. He postulated that presence of impurities instead of screw dislocations were responsible for the formation of etch pits[40], but not identify the specific type of impurity. Recent XRD and X-ray total scattering pair distribution function (XPDF) analyses have revealed that the gibbsite nanoplates synthesized in the same manner as this study consist of dense, truncated layers that cause local variations in the interlayer spacing [41, 42]. It appears that these interlayer defects can be accurately represented by flat Al13 aluminum hydroxide nanoclusters, which are nearly isostructural with the gibbsite layers present during the synthesis process. These nanoclusters become trapped as interlayer inclusions while the gibbsite nanoplates are growing[41]. We postulate these nanocluster inclusions respond to the formation of etch pits in gibbsite. When the nanoclusters become exposed due to dissolution, the strain disappears, and the dissolution pits will no longer continue to deepen. This behavior differs from the dissolution of gibbsite in an acidic environment, where in-situ AFM images from Peskleway et al. showed that screw dislocations in natural gibbsite were opened up in nitric acid solutions with a pH of -0.9.[32]

Anisotropic dissolution has been observed not only in the dissolution of hillocks but also in the growth of etch pits on the gibbsite basal plane, where the preferred directions change layer by layer. Figure 4f shows that the dissolution of deep pits also exhibits anisotropy. The retreat rates of the ($\bar{1}$00) step and adjacent steps on the basal plane are faster. Additionally, the morphology of dissolution pits in different layers is different, indicating that the anisotropy of dissolution activity varies among different layers. Figure 4g depicts the changes in step contours over time from the first to the fifth layer, where layers 3 to 5 represent the contours of different layers of the deep pit in Figure 4f. The step-orientation dependent dissolution rates of these layers were measured and plotted in Figure 4h. Layers 1 and 2 measure the rate of straight steps, while layers 3 to 4 correspond to the dissolution of etch pits. In different layers, the ($\bar{1}$00) step on the basal plane always dissolves faster compared to the (100) step (1.240 nm/min vs. 0.200 nm/min in the first layer). The dissolution rate of etch pits (layers 3-5) is lower than that of straight steps (layers 1-2), which can be attributed to the curvature effect and the limited diffusion of solute (ref: 43). The preferred direction of dissolution alters layer by layer, switching between the second quadrants (steps (010), ($\bar{1}$10), ($\bar{3}$10), and ($\bar{1}$00)) and the third quadrants (steps (0$\bar{1}$0), ($\bar{1}\bar{1}$0), ($\bar{3}\bar{1}$0), and ($\bar{1}$00)) of the crystallographic orientation of the basal plane. The dissolution rate spectra of adjacent layers exhibit mirror symmetry along the a-c plane, similar to the dissolution of hillocks in 0.1 M NaOH. We also observed consistent dissolution anisotropy in higher NaOH concentrations (0.5 M), as shown in Figure S5.

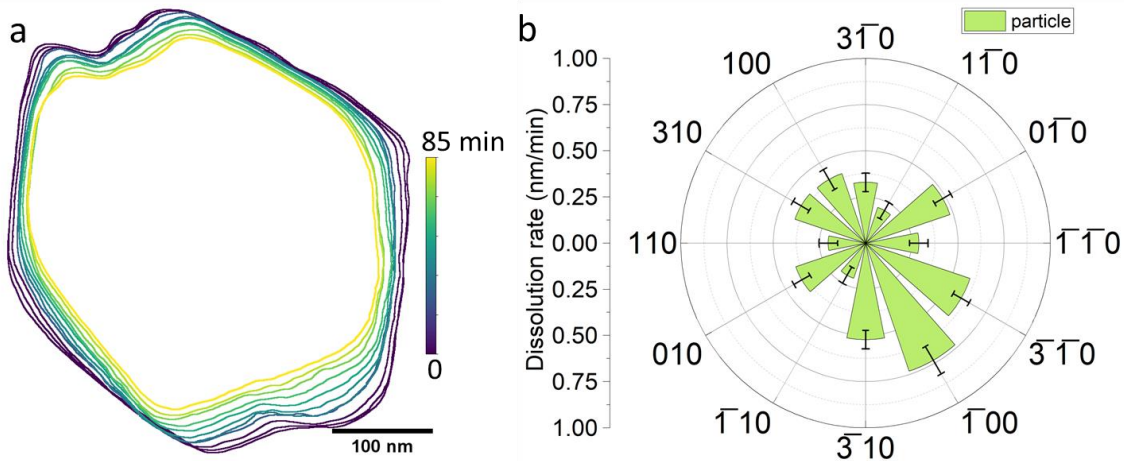

Figure 5. (a) The counter plot of gibbsite nanoplate dissolving in 0.2 M NaOH. (b) The dissolution rate of nanoplate edge as a function of edge orientation.

Despite the inconsistent anisotropic dissolution at gibbsite (001) and (002) plane, the periodic variation in preferred dissolution direction results in a more mirror-symmetrical dissolution rate-orientation relationship for the entire nanoplate along the a-c plane. Figure 5a illustrates the changes in particle outline over time in 0.2 M NaOH, while Figure 5b presents the measured dissolution rates. The dissolution rate of particle profiles can correspond to the dissolution rate of parallel steps, as opposed to the dissolution of a single step measured previously. Interestingly, among the zigzag types of parallel steps, the (-100) parallel steps exhibit the highest velocity and a comparable dissolution rate (0.732 ± 0.085 nm/min) with a single (-100) step on layer 1 (1.240 ± 0.538 nm/min). The four {110} parallel steps have a significantly slower dissolution rate (0.224 ± 0.037 nm/min). This rate is similar to the slower ($1\bar{1}0$) and (110) single steps on layer 1 (0.161 – 0.289 nm/min) but much lower than the faster ($\bar{1}\bar{1}0$) and ($\bar{1}10$) steps on layer 1 (0.587 ± 0.22 nm/min), demonstrating that the rate-limiting step of parallel step dissolution is controlled by the slower steps.

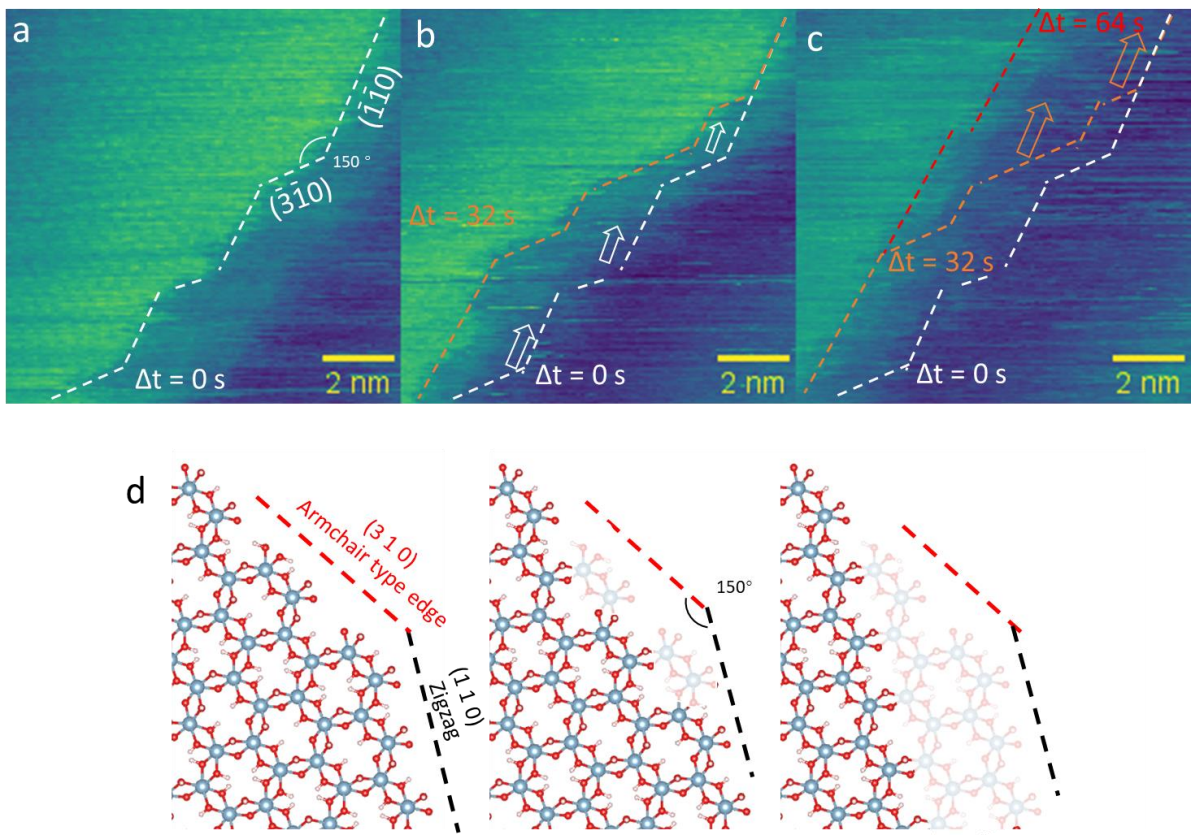

Figure 6. (a-c) High resolution AFM image shows the dissolution process of a gibbsite step transitioning from a kinked step consisted of ($\bar{1}\bar{1}0$) and ($\bar{3}\bar{1}0$) steps to a straight ($\bar{1}\bar{1}0$) step on the basal plane. (d) Schematic diagram of the retreat of ($\bar{3}\bar{1}0$) step.

**Mechanisms of anisotropic dissolution**

The continuous scanning of the oscillating AFM tip during measurements may potentially introduce artificial interference to the dissolution of mienral[44, 45]. To exclude the possibility that the anisotropic dissolution of gibbsite is induced by artifact, we tested the gibbsite dissolution rate in NaOH through batch experiments with intense stirring (Figure S6). The surface area normalized dissolution obtained from AFM measurements of single particles is very close to the rate measured in batch experiments at the same NaOH concentration (Table S1). This suggests that continuous scanning of AFM tip does not have a significant promoting effect on the dissolution rate.

Additionally, the fact that dissolution rates are the same under both static AFM measurements and intense stirring batch experiments suggests that gibbsite dissolution in the given NaOH concentrations is in the interfacial reaction-controlled regime.

Figure 5b demonstrates that armchair-type parallel steps ({310} and {010}) dissolve faster than zigzag-type {110} steps, suggesting greater stability for the latter. This observation is supported by high-resolution AFM studies of step dissolution in 0.1 M NaOH. Figure 6a-c depicts the dissolution process of a gibbsite step transitioning from a kinked to a straight step. The initial kinked step comprises zigzag-type ($\bar{1}\bar{1}0$) and armchair-type ($\bar{3}\bar{1}0$) steps, as seen in the atomic resolution AFM image. In Figure 6b, after 32 seconds, the ($\bar{3}\bar{1}0$) steps retreat parallel to the (110) step, while the initially straight (110) step remains undissolved. Complete retraction of the straight (110) step occurs only upon the continued retreat of the (310) step, as shown in Figure 6c.

These findings indicate that armchair-type steps dissolve faster than adjacent zigzag-type steps, and zigzag-type steps retreat is achieved by continuously dissolving newly formed or existing kinks consisting of armchair-type arrangements. Schematic diagram of step structure in Figure 6d reveals that the Al atoms in a straight step have four Al-OH-Al linkages to the crystal. For zigzag-type steps, dissolving one Al atom forms a double kink, where Al atoms are still connected to the crystal by four Al-OH-Al linkages. Further kink site extension requires overcoming a similar energy barrier. In contrast, armchair-type steps feature dimeric Al hydroxides. If one Al atom is lost, the other Al atom in the pair will have only two Al-OH-Al linkages with the crystal, making it more susceptible to dissolution within a shorter time. Consequently, armchair-type steps exhibit higher dissolution rates than zigzag-type step. Due to the interdependence of the two preceding and following aluminum atom dissolution events, the dissolution of the dimer can be considered the least independent process of gibbsite dissolution. Our results also support the previous speculations on the dissolution mechanism. Addai-Mensah et al. found that the gibbsite dissolving rate by NaOH depended upon the concentration of Al released in solution to the power of 2, and suggested that the dissolution process possibly involve rate-determining, $Al^{III}$-containing dimer formation and saturation at the gibbsite-NaOH solution interface[46]. In our study, such AlIII-containing dimers correspond to the armchair-type step.

However, this does not explain why the dissolution rate of the zigzag-type ($\bar{1}00$) step is comparable (Figure 2h and 4f) or even higher than that of the armchair-type ($\bar{3}10$) step (Figure 5b), suggesting that each type of Al-OH-Al linkage might have nonequivalent bond strength. Figure 7 shows the unit cell of gibbsite, which has six hydroxyl groups that create four types of Al-OH-Al pair linkages connecting adjacent aluminum atoms. These pairings are designated as O1/O2, O3/O3, O4/O4, and O5/O6. It is evident that the four types of $\mu_2$-O pairs have different orientations. Specifically, the O3/O3 and O4/O4 pairs are aligned in the <010> direction, whereas the O1/O2 pairs alternate between the <1-10> and <110> directions in adjacent layers. In contrast, the O5/O6 pairs are oriented in the opposite direction to the O1/O2 pairs, as shown in Figure 7. It is worth noting that ab initio molecular dynamics simulations performed on the basal surfaces of gibbsite found that the hydroxyl groups on the surface interact only weakly with the waters above and remain essentially undisturbed[47]. This implies that extracting the surface structure of the solid directly from the crystal structure remains representative of the interface in the solution environment

To compare the strength of each type of Al-OH-Al pair, we calculated the energy increase based on DFT by deliberately separating the gibbsite (001) plane into a surface trench and an Al(OH)3 stripe by breaking specific Al-OH-Al pairs in a periodic box (see more details of calculation in Supporting Information S2). The strength of the broken linkages is directly reflected in the energy increase by breaking per $\mu_2$-O pair (dE), which are tabulated in Table 1.

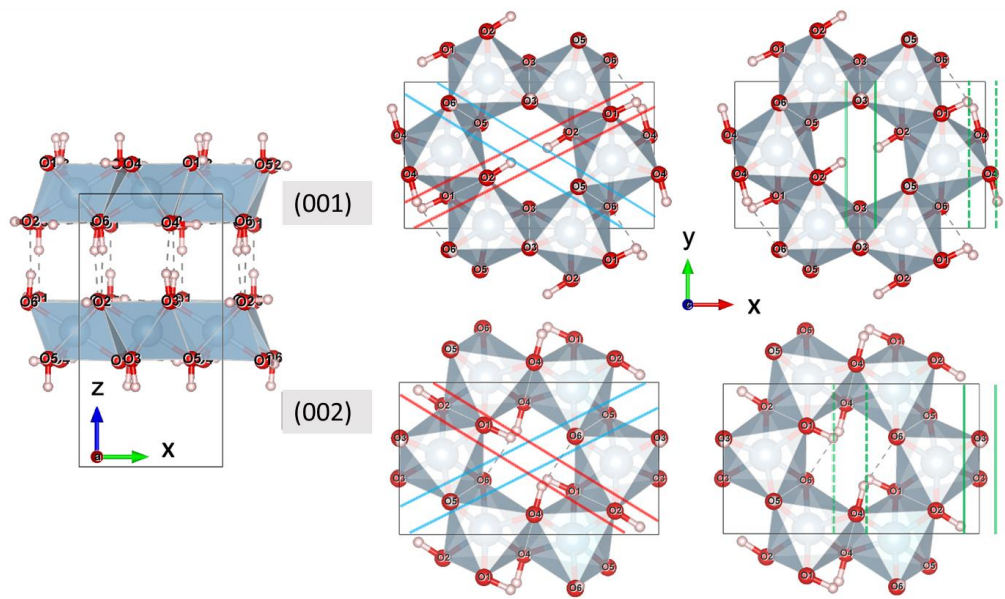

Figure 7. Gibbsite unit cell structure. The side view of the unit cell (a). The top view of the layer 1 (b) and layer 2 (c) with red, blue, green solid, and green dash lines indicating the O1/O2, O5/O6, O3/O3, and O4/O4 linkage pairs.

Table. 1 Calculated energy increasement after breaking per $\mu_2$-O pair along different directions

| Al-OH-Al link pair | The steps index be related | Relative energy increase (dE) | |
|---|---|---|---|
| | | meV | kcal/mol |
| O3/O3 | (100), ($\bar{1}$00) on (001)/(002) plane | 0 | 0.00 |
| O4/O4 | (100), ($\bar{1}$00) on (001)/(002) plane | 95.595 | 2.20 |
| O1/O2 | (1$\bar{1}$0), ($\bar{1}$10) on (001); (110),($\bar{1}$$\bar{1}$0) on (002) plane | 231.7275 | 5.34 |
| O5/O6 | (110),($\bar{1}$$\bar{1}$0) on (001); (1$\bar{1}$0), ($\bar{1}$10) on (002) plane | 191.6275 | 4.42 |

In Table 1, the lowest dE value which is from the O3/O3 pair is shifted to 0.0 and used as the reference point. Notably, the energy increasements for pairs aligned along the <010> direction, namely O3/O3 and O4/O4, are considerably lower than those in other directions. This observation is consistent with the much higher step retreat velocity (i.e., dissolution rate) of (100) and (-100) steps observed in nanoplates, compared to four {110} steps (Figure 5b). Conversely, the pairs O1/O2 and O5/O6 appear alternatively in the same direction (either <110> or <1-10>), depending on the layer, resulting in a similar retreat velocity for the corresponding parallel {110} steps (Figure 5b). Moreover, given the different strength of O1/O2 and O5/O6 pairs, the direction of preferential dissolving switches according to the same way of the alternation of direction of these pairs, namely has mirror symmetry respect to a-c plane. These results are consistent with the dissolution behavior of hillock and etch pits at basal plane (Figure 2 and 4). Our analysis thoroughly examines the forms of Al-OH-Al linkages and provides an estimate of their strength. Our DFT findings are consistent with experimental observations and explain the preferred direction of gibbsite nanoparticle dissolution. However, it is worth noting that our results do not account for the potential impact of the solution on the behavior of the $\mu_2$-O pairs. Also, our findings do not fully elucidate the substantial difference in dissolution rate observed in opposing directions within the same layer.

Although the differentiation of various linkage pairs is not able to elucidate the substantial difference in dissolution rate observed in opposing directions within the same layer, the asymmetry in the structure of the opposite directed steps could be a potential explanation. As shown in the schematic plot (Figure S7), we are considering a dissolution process that happens at the position noted by the dashed line and proceeds either to the left or the right. Our previous study has shown that an aluminate monomer dissolution is composed of a series of elementary bridge bond (Al-O) broken events[38]. Considering the same position (highlighted by purple oxygen atoms in Figure S7a), we assume that the easiest broken bond is the longest one as indicated in Figure S7b which is the same for either direction.

The first breakpoint is indicated in both Figure S7c and S7d by assuming the longest bond (1.99 Å) will be scissored first. Based on this assumption, when the dissolution proceeds to the left (Figure S7c), the cluster will carry a negative 0.5 e⁻ formal charge (inset of Figure S7c). However, when the dissolution proceeds to the right (Figure S7d), with the same bond broken at the same place, the detaching cluster would be assigned a positive 0.5 e⁻ charge which is completely opposite to the previous case. Meanwhile, no matter the dissolution proceeds in which direction, the solid edge would be negatively charged due to the high pH (> 13.0) circumstance, as the p$K_a$ value of the gibbsite edge is around 9.9[49]. So, the formal charge carried by the cluster after the presumptive first bridge Al-O bond is broken would either facilitate the cluster dissolution or hinder the cluster dissolution depending on the sign of the charge that the cluster is carrying due to the coulombic interaction.

**Conclusion**

In conclusion, by *in situ* AFM measurement at atomic-scale mapping, we reveal that the anisotropic dissolution behavior of gibbsite basal plane in NaOH solution, primarily governed by the interfacial atomic structure of gibbsite. The dissolution rates of hillocks, etch pits, and parallel steps on the basal plane display mirror symmetry along the a-c plane. The strength of Al-OH-Al pair linkages in varying directions is crucial in determining the preferred dissolution direction, and the anisotropy of dissolution differs from layer to layer. Our findings have significant implications for comprehending gibbsite dissolution mechanisms in alkaline environments and for developing more precise models to predict and regulate dissolution rates and pathways. This research also encourages further exploration into the anisotropic dissolution of other layered materials, such as micas, clays, and layered double hydroxides, as well as the impact of step orientation and interfacial structure on their growth/dissolution behavior. Additionally, the DFT method employed in this study to assess the strength of Al-OH-Al pair linkages could potentially be applied to other materials with analogous structures for a better understanding of their dissolution mechanisms.

**Materials and Methods**

**Synthesis of Gibbsite Nanoplates**

Samples were synthesized using a hydrothermal method, following established procedures[36]. Specifically, 0.25 M aluminum (Al) (Al(NO3)3·9H2O, ≥98%, Sigma-Aldrich) was dissolved in deionized water (18.20 MΩ/cm) with stirring, and the pH was adjusted to approximately 5.0 by adding a 3 M sodium hydroxide (NaOH) (≥98%, Sigma-Aldrich) aqueous solution. The resulting solution was continuously stirred for 1 hour, then centrifuged to collect gel-like precipitates. The gel was washed three times with deionized water and dispersed into water to create 0.25 M suspensions based on Al ions. Subsequently, 16 mL of the gel solution was transferred into a 20 mL Teflon vessel, which was then sealed within a Parr bomb and heated at 80 °C for 5 days. Finally, the resulting white product was collected by centrifugation, washed three times with DI water, and dried at 50 °C overnight.

**Solids Characterization**

X-ray diffraction (XRD) patterns were obtained utilizing a Philips X'pert Multi-Purpose Diffractometer (MPD, PANAlytical, Almelo, The Netherlands) outfitted with a stationary Cu anode running at 50 kV and 40 mA. The XRD patterns were documented within a 5–80° 2θ range. Phase identification was accomplished through the application of JADE 9.5.1 (Materials Data Inc.) and the 2012 PDF4+ database from the International Center for Diffraction Data (ICDD). Morphology characterization was performed using a Helios NanoLab 600i SEM (FEI, Hillsboro, OR). A carbon coater was used to coat a thin layer carbon (∼5 nm) on all samples prior to analysis to improve the conductivity and imaging. Transmission electron microscopy (TEM) samples were prepared by dispersing in deionized water through 5-minute sonication and subsequently drop-casting the resulting mixture onto copper grids (Lacey Carbon, 300 mesh, Ted Pella, Inc.). The grids were then allowed to dry under ambient conditions. Cs-corrected field emission transmission electron microscope (FEI, Hillsboro, OR) was employed to carry out the TEM analysis, with all samples being imaged at an acceleration voltage of 300 kV.

**Atomic Force Microscopy**

The Cypher VRS AFM was used to perform all AFM measurements at room temperature. The FastScanB AFM tip with a spring constant of 1.8 N/m was utilized in this study. To evaluate the dissolution kinetics of gibbsite at various NaOH concentrations (0.1-0.5 M), the AFM tip was brought to the sample surface using the amplitude mode, and the scan rate for dissolution kinetics images was set to 2 Hz. The scan was conducted for over 2 hours, resulting in more than 30 images to calculate the dissolution kinetics. Additionally, high-resolution images of gibbsite were obtained by scanning the AFM tip over a 20 × 20 nm$^2$ area at a scan rate of over 8 Hz.

**Dissolution Experiments**

The batch dissolution experiments were performed by dispersing 0.43g of gibbsite nanoplates into 100 mL 0.1, 0.2, and 0.5 M NaOH solution in polyethylene bottles. The suspensions were stirred at 400 rpm by magnetic stirrers. We periodically took samples of the solution at different time intervals and filtered each sample through a 0.2-micron filter membrane to remove any solids or particles. The filtered solution was then acidified using 2M HCl and diluted by 2% $HNO_3$. We used Inductively Coupled Plasma Optical Emission spectroscopy (ICP-OES) analysis to determine the concentration of aluminum (Al) in the filtered and acidified solution.

**Density Functional Theory Calculation**

Density functional theory (DFT) simulations are carried out with Vienna Ab initio Simulation Package (VASP) code[50] with generalized gradient approximation (GGA) in the formulation of Perdew-Burke-Ernzerhof (PBE)[51] along with the projector augmented wave (PAW) method[52]. The energy cutoff for the plane wave basis sets is set to 400 eV. The slab systems are fully optimized with force criteria of 0.05 eV/Å. The K-points density, which determines the spacing between k-points in the Brillouin zone, is controlled by the KSPACING keyword. A value of 0.5 Å-1 is used for geometry optimization, and 0.2 Å-1 is used for static calculations. The convergence criterion for the total energy is set to 10-7 eV for static calculations, which are used for energy comparison purposes. More detail information about the estimation of bonding strength of Al-OH-Al linkage pair in the gibbsite basal plane is described in Supporting information S2.

**Acknowledgements**

This research was supported by Interfacial Dynamics in Radioactive Environments and Materials (IDREAM), an Energy Frontier Research Center funded by the U.S. Department of Energy, Office of Science, Basic Energy Sciences (FWP 68932). A portion of the research was performed in the Environmental Molecular Sciences Laboratory (EMSL), a national scientific user facility sponsored by the DOE Biological and Environmental Research program and located at PNNL. PNNL is a multiprogram national laboratory operated for DOE by Battelle Memorial Institute under Contract No. DE‐AC05‐76RL0‐1830. The authors acknowledge fruitful discussions with Dr. Shuai Zhang at University of Washington and Shawn Riechers at PNNL We also thank PNNL scientist Charles T. Resch for help with the ICP-OES measurements.